\documentclass[aps,prb,twocolumn,showpacs]{revtex4}
\usepackage{graphicx}

\newcommand{\cm}{\ensuremath{\mbox{cm}^{-1}}}

\begin{document}

\title{Central mode and soft mode behavior in $\mathrm{PbMg_{1/3}Nb_{2/3}O_3}$ relaxor ferroelectric}

\author{S. Kamba, M. Kempa, V. Bovtun, and J.
Petzelt}
 \affiliation{Institute of Physics, Academy of Sciences of the Czech
Republic, Na Slovance 2, 182 21 Prague 8, Czech Republic}
\author{K. Brinkman and N. Setter}
\affiliation{Ceramics Laboratory, EPFL,Swiss Federal Institute of Technology,
1015 Lausanne, Switzerland}

\date{\today}

\begin{abstract}
The relaxor ferroelectric $\mathrm{PbMg_{1/3}Nb_{2/3}O_3}$ was investigated by means of
broad-band dielectric and Fourier Transform Infrared (FTIR) transmission spectroscopy in
the frequency range from 1 MHz to 15 THz at temperatures between 20 and 900 K using PMN
films on infrared transparent sapphire substrates. While thin film relaxors display
reduced dielectric permittivity at low frequencies, their high frequency intrinsic or
lattice response is shown to be the same as single crystal/ceramic specemins. It was
observed that in contrast to the results of inelastic neutron scattering, the optic soft
mode was underdamped at all temperatures. On heating, the TO1 soft phonon followed the
Cochran law with an extrapolated critical temperature equal to the Burns temperature of
670\,K and softened down to 50\,cm$^{-1}$. Above 450\,K the soft mode frequency leveled
off and slightly increased above the Burns temperature. A central mode, describing the
dynamics of polar nanoclusters appeared below the Burns temperature at frequencies near
the optic soft mode and dramatically slowed down below 1\,MHz on cooling below room
temperature. It broadened on cooling, giving rise to frequency independent losses in
microwave and lower frequency range below the freezing temperature of 200 K. In addition,
a new heavily damped mode appeared in the FTIR spectra below the soft mode frequency at
room temperature and below. The origin of this mode as well as the discrepancy between
the soft mode damping in neutron and infrared spectra is discussed.
\end{abstract}

\maketitle

\section{Introduction}

Lead magnesium niobate PbMg$_{1/3}$Nb$_{2/3}$O$_3$~(PMN) was initially studied by
Smolenskii $et$ $al.$ \cite{Smolenskii58, Smolenskii60} in the end of 1950's, and since
that time it has been considered as a model system for relaxor ferroelectrics (RFE). The
interest in this topics was recently enhanced after the paper of Park and Shrout
\cite{Park97} who showed that PbMg$_{1/3}$Nb$_{2/3}$O$_3$-PbTiO$_{3}$ and
PbZn$_{1/3}$Nb$_{2/3}$O$_3$-PbTiO$_{3}$ mixed crystals exhibit utrahigh strain and
piezoelectric behavior. The model system, PMN shows high and broad maxima of real and
imaginary parts of the complex permittivity $\varepsilon^*$(T) = $\varepsilon'$(T)
-i$\varepsilon''$(T) which shift with increasing measuring frequency to higher
temperatures. The dielectric anomaly is not accompanied by any phase transition to polar
state, and the structure remains cubic down to liquid He temperatures.\cite{Mathan}
Nowadays it is generally accepted that the strong dielectric dispersion, which occurs in
RFE in the broad spectral range from GHz to milliherz range\cite{Bovtun04,Colla93} at
temperatures around and below that of $\varepsilon'$(T) maximum T$_{max}$, is a
consequence of the dynamics of polar nano-clusters, which appear several hundred degrees
above T$_{max}$, at so called Burns temperature T$_{d}$.\cite{Burns83} Although many
papers have been devoted to RFE (see e.g. reviews by Cross\cite{Cross87}, Ye\cite{Ye98},
Samara\cite{Samara01} and Kamba\cite{Kamba04}), the quantitative description of the
complex dielectric behavior of RFE is still not satisfactory.

The work presently available on thin film relaxors show that the permittivity is reduced
by an order of magnitude as compared to single crystal or ceramic
specimens\cite{Zian,Tyunina}. The commonly reported mechanisms for this reduction in thin
films are grain size \cite{Papet}, clamping of the film by the substrate \cite{Catalan},
and passive or surface layer effects\cite{Park}. However these arguments rely on the
assumption that the intrinsic or lattice response of the material is the same in thin
film and "bulk" ceramic/single crystal form, and that somehow the extrinsic response from
polar clusters/regions is impacted due to grain size, stress or pinning by processing
induced defects. In fact, there has not been any reports on the model PMN relaxor system
in thin films form which describe the lattice dynamics and how they compare to single
crystal specimens. In the course of this paper we will show that indeed, the high
frequency lattice response (in THz range) is the same for PMN in thin film and single
crystal specimens, thus providing an important "basis" on which to judge the origins of
the reported reduced dielectric response in thin films. In addition, thin film specimens
are more convenient for use in transmission FTIR measurements, so that along with
studying the material in its thin films form, we may comment on the intrinsic nature of
the material itself as compared to other measurement techniques.

Naberezhnov $et$ $al.$ performed the first study of the lattice dynamics of PMN single
crystal by means of inelastic neutron scattering (INS)\cite{Naberezhnov99}. They observed
a transverse optic (TO1) phonon branch which softens on cooling to T$_{d}\approx$ 620\,K.
Below T$_{d}$ the softening ceased and a strong central peak appeared. Gehring $et$ $al.$
discovered that below T$_{d}$ the TO1 branch appeared to dive into the transverse
acoustic (TA) branch at a specific wavevector $q_{WF}\approx 0.20$\,$\AA$ and no TO1
phonon could be resolved from the INS spectra at lower
$q$.\cite{Gehring00a,Gehring00b,Gehring01b} Gehring $et$ $al.$ named this effect the
"waterfall", and explained it by the presence of polar clusters below T$_{d}$. According
to their arguments, when the wavelength of the TO1 mode becomes comparable to the size of
the polar clusters, it cannot propagate and thus becomes overdamped. Simultaneously, it
was observed that the damping of TA mode remarkably increased below
T$_{d}$.\cite{Naberezhnov99,Wakimoto02a} The waterfall effect is not specific only for
PMN, but has been observed in other relaxors or relaxor based crystals like
$\mathrm{PbZn_{1/3}Nb_{2/3}O_3}$, $\mathrm{PbMg_{1/3}Nb_{2/3}O_3-PbTiO_{3}}$ and
$\mathrm{PbZn_{1/3}Nb_{2/3}O_3-PbTiO_{3}}$.\cite{Gehring01a,Tomeno01,Koo02,Gehring00a}
The line shape of TA and TO1 phonon peaks in INS experiment were successfully explained
by the coupling of both modes with the assumption that the TO linewidth (damping) is
wave-vector and temperature dependent.\cite{Shirane01,Wakimoto02b} Hlinka {\itshape et
al.} repeated the INS measurements of PZN-PT and discovered that the critical waterfall
wave vector depends on the Brillouin zone, where the experiment is
performed.\cite{Hlinka03} Therefore, they claimed that the waterfall effect is not
connected with phonon scattering on polar nanoclusters but could be explained as apparent
effect in the framework of a model of bilinearly coupled harmonic oscillators
representing the acoustic and heavily damped optic phonon branches and the fact that
there are different dynamical structure factors in different Brillouin zones.

Wakimoto {\itshape et al.} cooled down the PMN crystal and discovered that the soft TO1
mode recovers (i.e. underdamps) in the INS spectra below 220 K and its frequency hardens
according to Cochran law with decreasing temperature.\cite{Wakimoto02a,Wakimoto02b} This
result was rather surprising because such behavior is typical for displacive feroelectric
transitions below T$_{c}$, but PMN remains paraelectric down to liquid He
temperatures.\cite{Mathan} It means, that the soft mode behavior provides evidence about
the ferroelectric order in polar clusters. On the other hand, it is interesting to note
that 220\,K coincides with the ferroelectric phase transition temperature T$_{c}$ in PMN
cooled under an electrical bias field.\cite{Ye93} The question remains, how does the TO1
zone center soft mode (SM) behave between T$_{c}$ and T$_{d}$, where it is not resolved
in the INS spectra? Direct experimental data are missing, but Stock {\itshape et al.}
extrapolated the measured phonon dispersion curves of the soft TO1 branch from high
{\itshape q} to the zone center and predicted that on heating the TO1 mode softens only
to 400\,K after which it again slightly hardens.\cite{Stock04}. In this report, we have
used Fourier transform infrared (FTIR) transmission techniques to experimentally examine
the response of the material in this temperature range where the SM was unobserved by
neutron scattering methods.

Kamba {\itshape at al.} measured FTIR reflectivity of PMN-29\%PT and PZN-8\%PT single
crystals, and in contrast to INS data, they have seen the lowest frequency TO1 phonon
underdamped at all temperatures between 10 and 530\,K.\cite{Kamba03} This discrepancy was
explained by the different {\itshape q} vectors probed in infrared (IR) and neutron
experiments. The IR probe couples with very long-wavelength phonons
($q\approx$10$^{-5}$\,\AA$^{-1}$) which see the homogeneous medium averaged over many
nanoclusters, whereas the neutron probe couples with phonons whose wavelength is
comparable to the nanocluster size ($q\ge$10$^{-2}$\AA$^{-1}$).\cite{Kamba03} Recently
Kamba {\itshape at al.} also investigated PMN single crystal by means of FTIR reflection
spectroscopy and observed an underdamped TO1 SM at all temperatures below 300\,K which
obeyed the Cochran law.\cite{Bovtun04,Prosandeev04} Unfortunately they were not able to
evaluate unambiguously FTIR data above room temperature, because the reflection band from
the TO1 mode overlaps with that of the central mode (CM). Therefore we have decided to
perform more sensitive FTIR transmission measurements on a thin PMN film.

\section{Experimental}

PMN single crystals are opaque in the far IR range due to the strong polar mode
absorption down to thicknesses of the order of 1 $\mu$m. Therefore we prepared PMN thin
films by chemical solution deposition on sapphire substrate, well transparent in the
far-IR range. The PMN film had thickness of 500\,nm, the plane-parallel sapphire
substrate had the size of 6x8\,mm, a thickness 490\,$\mu$m and both faces were optically
polished. The film was polycrystalline with a predominately (111) out of plane
orientation and a nominal grain size of 60 nm.

In-plane dielectric response of the film was measured using a planar capacitor with standard photolithographically defined gaps of 10 microns between Cu/Cr electrodes sputter deposited on the surface of the PMN film.  the capacitance and loss tangent of the electrode/film/substrate stack was measured using a HP4284A impedance analyzer at frequencies from 10 kHz to 1MHz at temperatures from 375 to 175 K.  The permittivity of the film was calculated from the electrode/film/substrate stack capacitance according to Vendik \cite{Vendik} with an estimated error of +/- 10\%.

The unpolarized FTIR transmission spectra were taken using a FTIR spectrometer Bruker IFS
113v at temperatures between 20 and 900\,K with the resolution of 0.5\,\cm. A helium
cooled Si bolometer operating at 1.5\,K was used as a detector, while an Optistat CF
cryostat with polyethylene windows was used for cooling, and a commercial
high-temperature cell SPECAC P/N 5850 was used for the heating. The investigated spectral
range was determined by the transparency of the sapphire substrate; at 20~K up to
450\,\cm\ (0.3-15 THz), at 900\,K the sample was opaque above 190\,\cm.

High-frequency and MW dielectric spectra were obtained on PMN single crystal using
Agilent HP 4291B impedance analyzer (1MHz-1.8\,GHz; 100-520\,K) and by waveguide
resonance and non-resonance method using fixed-frequency generators (7-74\,GHz, 100-800
\,K).\cite{Bovtun97}

\section{Results and discussion}

\begin{figure}
\centerline{\includegraphics[width= 80mm, clip=true] {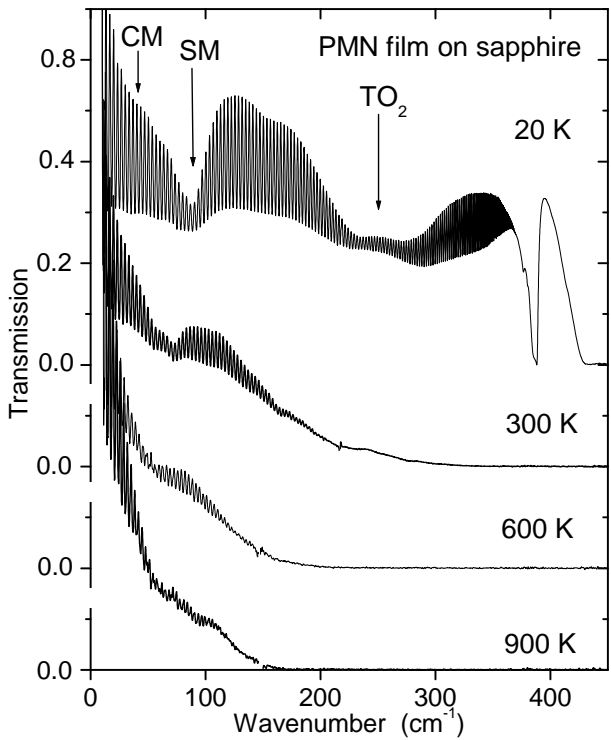}} \caption{FTIR
transmission spectra of the 500\,nm thick PMN film deposited on sapphire substrate
(490\,$\mu$m). Frequencies of the central mode (CM), TO1 soft mode (SM) and TO2 phonon
are marked. Absorption peak near 380\,\cm\ is the phonon peak from the sapphire.
\label{fig:trans}}
\end{figure}

An example of the FTIR transmission spectra of the PMN thin film on a sapphire substrate
at selected temperatures is shown in Fig.~\ref{fig:trans}. The dense oscillations in the
spectra are due to interferences in the substrate, while broad minima correspond to
frequencies of polar phonons. The transmission decreases on heating, mainly due to the
increase in multi-phonon absorption of the sapphire substrate. Unusual increase of
transmission below 50\,\cm\ was observed on heating above 700\,K, which will be explained
below by disappearance of the CM above T$_{d}$.

The spectra of a bare substrate and the PMN film on the substrate were determined for
each temperature studied. For a given temperature, the transmission spectrum of the bare
substrate was first fitted with a sum of harmonic oscillators using Fresnel formulae for
coherent transmission of a plane-parallel sample (i.e. taking into account the
interference effects).\cite{Born} The resulting parameters of sapphire were then used for
the fit of the PMN/sapphire two-layer system. The complex transmittance of the two-layer
system was computed by the transfer matrix formalism method including interference
effects.\cite{Heavens}

\begin{figure}
\centerline{\includegraphics[width= 70mm, clip=true] {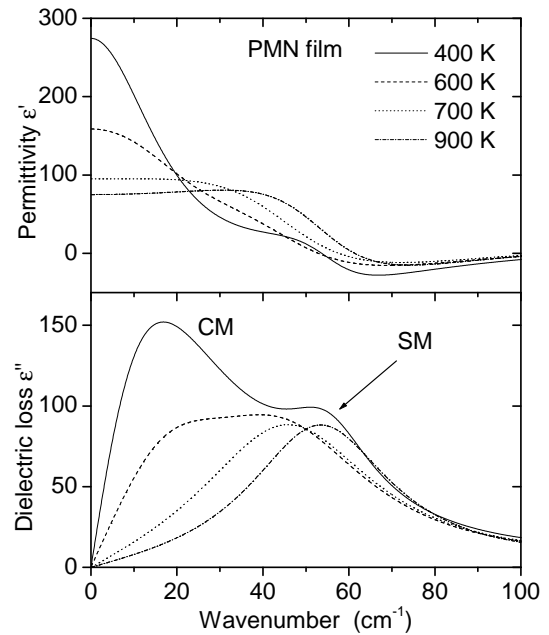}} \caption{Complex
dielectric spectra of PMN obtained from the fit of high-temperature FTIR transmission
spectra. \label{fig:epsHT}}
\end{figure}

\begin{figure}
\centerline{\includegraphics[width= 70mm, clip=true] {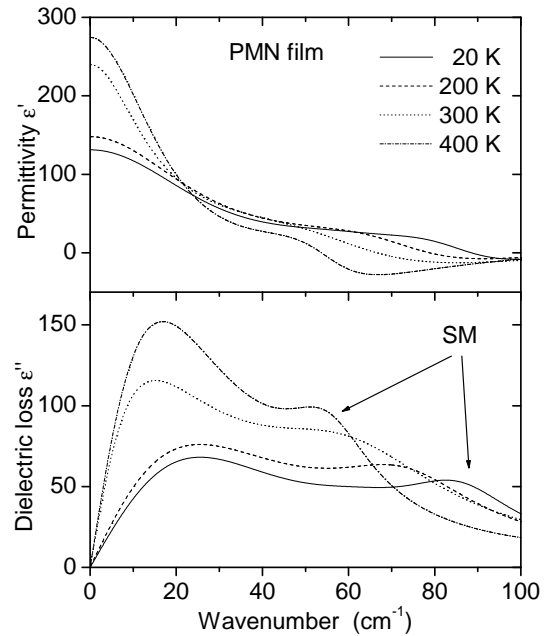}} \caption{Low-temperature
complex dielectric spectra of PMN obtained from the fit of FTIR transmission spectra.
\label{fig:epsLT}}
\end{figure}

The complex dielectric function expressed as the sum of damped quasi-harmonic oscillators
has the form\cite{Petzelt}
\begin{equation}
\label{eps3p}
 \varepsilon^*(\omega) =
\varepsilon'(\omega)-\textrm{i} \varepsilon''(\omega)
 = \varepsilon_{\infty} + \sum_{j=1}^{n}
\frac{\Delta\varepsilon_{j}\omega_{j}^{2}} {\omega_{j}^{2} -
\omega^2+\textrm{i}\omega\gamma_{j}}
\end{equation}
where $\omega_{j}$, $\gamma_{j}$ and $\Delta\varepsilon_{j}$ denote the frequency,
damping and contribution to the static permittivity of the j-th polar mode, respectively.
$\varepsilon_{\infty}$ describes the high-frequency permittivity originating from the
electronic polarization and from polar phonons above the spectral range studied. The mode
parameters at 20\,K are listed in Table~\ref{tab:IRmodes}. TO2 mode near 230\,\cm\ is not
seen at high temperatures due to opacity of the sapphire substrate, but at lower
temperature it can be well resolved and finally even its splitting is seen into three
components due to breaking of symmetry in polar clusters. We note that the factor group
analysis and the mode activities in the spectra of various phases of complex perovskite
ferroelectrics were published in Ref. \cite{Kamba03}

\begin{table}
\caption{Parameters of polar modes in PMN thin film obtained from the fit of FTIR
transmission spectrum at 20 K. Frequencies $\omega_{j}$ and dampings $\gamma_{j}$ are in
\ensuremath{\mbox{cm}^{-1}}, $\Delta \varepsilon_{i}$ is dimensionless,
$\varepsilon_{\infty}$=6.0. We note that the list of polar mode parameters is not
complete, because we do not see the phonons above 400\,\cm\ observed in single
crystal\cite{Prosandeev04} due to opacity of the substrate at high frequencies.}
\begin{tabular}{|l c c c |}\hline
  No&$\hspace{0.2cm} \omega_{j} \hspace{0.2cm}$&\hspace{0.2cm}
   $\Delta\varepsilon_{j}$\hspace{0.2cm}&\hspace{0.2cm} $\gamma_{j}$ \hspace{0.2cm}
 \\ \hline \hline
 CM&39.0&95.0&73.0\\
  1 &64.6&6.0&37.0\\
  2 (SM)&87.0&13.0&34.0\\
  3&153.0&1.5&56.0\\
  4&239.0&7.7&113.0\\
  5&276.0&0.9&42.2\\
  6&298.0&0.3&36.0\\
  7&347.0&0.3&29.0\\
  \hline
\end{tabular}
\label{tab:IRmodes}
\end{table}

\begin{figure}
\centerline{\includegraphics[width=80mm,clip=true] {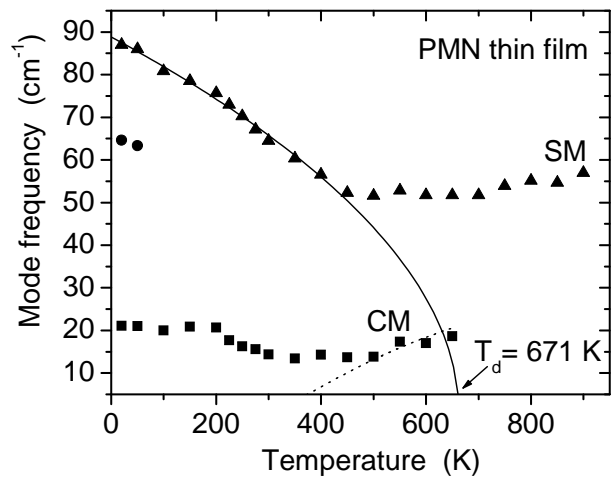}} \caption{Temperature
dependences of the SM and CM frequencies. CM is overdamped, therefore the frequency of
loss maximum corresponding to $\omega_{CM}^2/\gamma_{CM}$ is plotted. The Cochran fit of
the SM is shown by solid line, slowing down of the CM is schematically shown by dashed
line. \label{fig:SM-T}}
\end{figure}

The $\varepsilon^{*}(\omega)$ spectra calculated from the fit of the transmission spectra
of PMN/sapphire above and below 400\,K are shown in Figs.~\ref{fig:epsHT} and
~\ref{fig:epsLT}, respectively. For our discussion it is instructive to look at the
dielectric loss $\varepsilon''(\omega)$ spectra, because the frequency of loss maxima
characterize the frequencies of excitations, in our case the SM and CM frequencies even
in the case of their overdamping. Their temperature dependences are shown in Fig.
~\ref{fig:SM-T}. The TO1 mode slightly softens from 60\,\cm\ (900\,K) on cooling, but below
450\,K it starts to harden and follows  the Cochran law
\begin{equation}
\label{Cochran}
  \omega_{SM}^{2}=A(T_{d}-T)
\end{equation}
with the extrapolated critical temperature T$_{d}$=(671 $\pm$ 10)\,K and
A=(11.9$\pm$0.2)\,K$^{-1}$. This temperature is close to the Burns temperature 620\,K as
reported for PMN single crystals.\cite{Ye98} The SM frequency in the thin PMN film shown
in Fig.~\ref{fig:SM-T} is the same as in the bulk sample measured by means of FTIR
reflectivity below 300\,K\cite{Bovtun04,Prosandeev04} and INS spectroscopy obtained below
T$_{c}$ and above T$_{d}$.\cite{Wakimoto02a} It shows that the lattice vibrations are not
appreciably influenced by the size effect and by the possible strain in the film, and that the \emph{intrinsic or lattice response of relaxors is nominally the same in thin films as in single crystals.}

Note the appearance of a new mode near 65\,\cm\ which is seen below 50\,K. It occurs due
to the local rhombohedral symmetry of polar clusters. The activation of this mode in the
FTIR spectra is allowed in the whole temperature range below T$_{d}$ but it is only
distinguished as the shoulder near SM frequency at the lowest observed temperatures. This
is due to the strong absorption from the overdamped mode near 20\,\cm\ at high
temperatures and low damping of both SM and 65\,\cm\ mode at 20\,K.

It is important to note that the damping of the SM is only slightly temperature
dependent. It increases from 37\,cm\ (at 20\,K) to 50\,\cm\ (at 300\,K) and at higher
temperature it remains temperature independent within the accuracy of our fits. These
results show that the SM is \emph{underdamped} in the whole investigated temperature
range as predicted by Stock \cite{Stock04} for measurments in the Brillioun zone center,
and in direct contradiction with the INS spectra where \emph{overdamped} zone center TO1
SM is seen between T$_{c}$ and T$_{d}$.\cite{Wakimoto02a}. It is noted that the
definition overdamping of the phonon mode occurs when the ratio of the damping
$\gamma_{j}$ and frequency $\omega_{j}$ becomes higher than 2.\cite{Petzelt} In this case
the frequency of the $\varepsilon''(\omega)$ maximum does not correspond to $\omega_{j}$
but rather to $\omega_{j}^2/\gamma_{j}$. The discrepancy between the SM damping in INS
and FTIR spectra will be explained below.

\begin{figure}
\centerline{\includegraphics[width=80mm,clip=true] {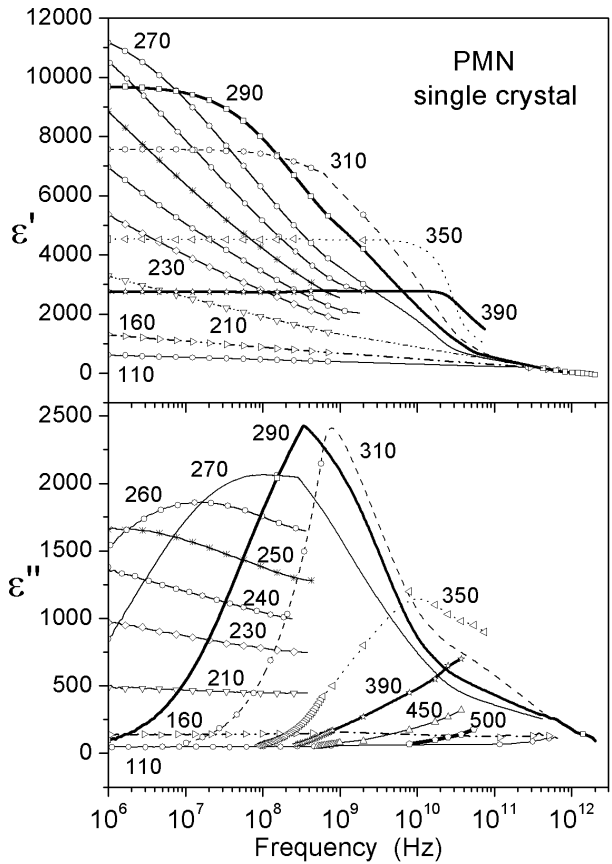}} \caption{Frequency
dependence of complex permittivity in high-frequency and MW frequency range at various
temperatures. \label{fig:eps-HF}}
\end{figure}

Below the Burns temperature T$_{d}\cong$670\,K the FTIR transmission remarkably decreases
below 40\,\cm\ due to a new overdamped excitation appearing below the SM frequency. It is
responsible for the the increase of the low-frequency $\varepsilon''$ (and
$\varepsilon'$) on reducing the temperature to 400\,K. We believe that this relaxational
excitation originates from the dynamics of polar clusters in PMN and can be called CM in
analogy with the INS scattering experiments. The CM frequency rapidly decreases on
cooling, and the fit to the overdamped oscillator is only approximate, especially at low
temperatures when the relaxation frequency lies below our frequency range. To get insight
into this range, we have performed dielectric measurements between 1\,MHz and 56\,GHz
down to 100\,K on single crystal PMN specimens (see Fig. ~\ref{fig:eps-HF}). One can see
the CM near 10\,GHz at 350\,K while it slows down to 10\,MHz at 260\,K. Simultaneously
the relaxation broadens. It follows from our previous dielectric measurements that the
mean relaxation frequency $\omega_{R}$ obeys the Vogel-Fulcher law
\begin{equation}
\label{Vogel}
  \omega_{R}=\omega_{\infty} \exp\frac{-E_{a}}{T-T_{VF}}.
\end{equation}
with the freezing temperature T$_{VF}$=200\,K, activation energy E$_{a}\sim$800\,K and
high temperature limit of the relaxation frequency $\omega_{\infty}\sim$ 5.7\,THz
(=190\,\cm).\cite{Bovtun04} At lower temperatures the loss maximum becomes so broad that
only frequency independent dielectric losses can be seen. It can be fitted by a uniform
distribution of relaxation frequencies, much broader than the experimental frequency
window.\cite{Kamba00} We suggest that this is due to the influence of random fields on
the distribution of activation energies for breathing of polar clusters.\cite{Bovtun04}

\begin{figure}
\centerline{\includegraphics[width=80mm,clip=true] {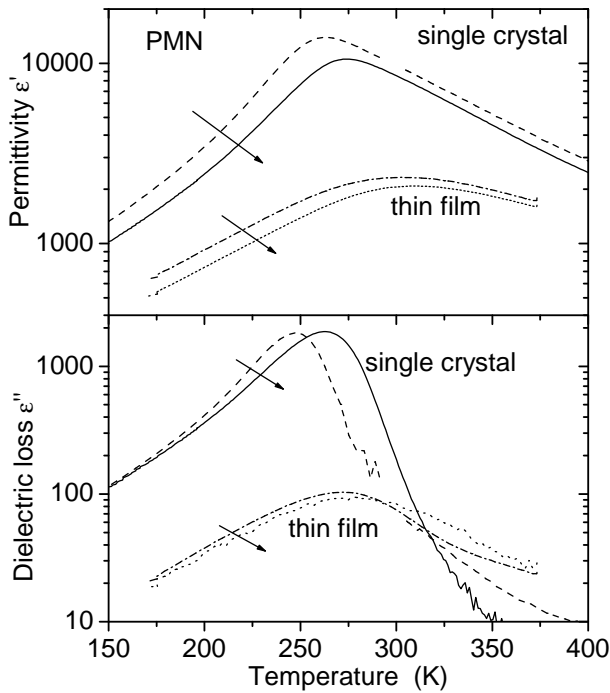}} \caption{Comparison of
temperature dependence of complex permittivity of PMN single crystal (from Ref.
\cite{Bovtun04}) and PMN thin film obtained at 10\,kHz and 1\,MHz (arrows mark the
increase of frequency). Dielectric response of the thin film was obtained in plane of the
film.\label{fig:comparison}}
\end{figure}

We measured also the in-plane dielectric response of PMN thin film on sapphire between 10
kHz and 1 MHz. Comparison of dielectric data obtained on single crystal and thin film is
shown in Fig. ~\ref{fig:comparison}. Relaxor behavior was observed also in the thin film,
only the dielectric maximum was approximately five times lower than in single
crystal\cite{Bovtun04} and T$_{max}$ was roughly by 50\,K higher, similar to recent reports
for PMN films measured in the "out of plane" or parallel plate capacitor configuration \cite{Zian}.
As we previously mentioned, the phonon behavior seems to be the same in single crystal and thin film, with the
phonon contribution to the static $\varepsilon$' being less than 100 in both cases. It seems
that the CM behavior is also qualitatively the same, however the dielectric strength of the
CM is smaller in the thin film. It is probably due to influence of size effect on the
dynamics of polar clusters known also in ceramics with different grain size.\cite{Papet}.
However, measurements of epitaxial films and films with different stress conditions also
showed reduced dielectric response \cite{Nagy} as compared to single crystal materials so
a detailed study of the impact of external factors leading to a reduced dielectric response
in thin films is urgently needed in order to address these questions.  Nevertheless, in both
single crystal and thin film at T$_{d}$ the CM appears near the SM
frequency and its mean relaxation frequency dramatically slows down on cooling to
T$_{VF}$. CM slows down faster in thin film, therefore T$_{max}$ is shifted to higher
value than in the single crystal. The huge temperature dependence of the CM and its
broadening is responsible for diffuse and frequency dependent maxima of
$\varepsilon^{*}$(T) in both single crystal and thin film. Egami\cite{Egami} has shown
that large displacements of Pb ions play the main role in creation of the local dipole
moment in nano-clusters below T$_{d}$. Therefore we can roughly assign our CM to strongly
anharmonic hopping of Pb ions.

From our FTIR spectra the CM has the highest dielectric strength $\Delta\varepsilon_{CM}$
at 400 K. At lower temperatures its frequency lies below our frequency range (see dashed
line in Fig.~\ref{fig:SM-T}) therefore $\Delta\varepsilon_{CM}$ apparently decreases on
cooling, although it should be highest at T$_{max}$. Nevertheless, some broad excitation
remains in the spectra down to 20\,K where its loss maximum stabilizes near the frequency
$\omega_{CM}^2/\gamma_{CM}$=20\,\cm. Its possible assignment is the activation of
low-energy phonons from the whole Brillouin zone in FTIR spectra (in this case phonons
from acoustic branch) as a consequence of breaking the translation symmetry due to
chemical disorder at the perovskite B sites and/or a locally doubled unit cell in polar
clusters.

Vakhrushev and Shapiro\cite{Vakhrushev} reported that not the TO1 SM but another
quasi-optic mode near 3.5\,meV exhibits softening on cooling to T$_{d}$. Their results
are not generally accepted because other authors\cite{Gehring01b,Wakimoto02a} including
this report have seen softening of the TO1 mode. Nevertheless, it is true that the
temperature behavior of the SM does not explain Curie-Weiss behavior of the static
permittivity above T$_{d}$\cite{Wakimoto02b} and another soft polar excitation is needed.
If the CM exists and hardens above T$_{d}$, it cannot be connected with dynamics of polar
clusters and its intensity should be very low so that it lies below our detection limit.
On the other hand, another SM could lie above the TO1 frequency. Unfortunately, we could
not study in detail the temperature dependence of TO2 mode near and above T$_{d}$ due to
opacity of the sapphire substrate. However we may comment on the behavior of the TO1 mode
and its apparent contradiction with previous reports using neutron scattering techniques.

From the results discussed herein, we suggest a plausible explanation for the discrepancy
between the SM mode observed by INS and FTIR techniques.  INS spectroscopy has a lower energy and $q$
 resolution than FTIR spectroscopy, therefore INS may
not resolve the splitting of the SM below T$_{d}$ into SM and CM, as it is seen in
Fig.~\ref{fig:epsHT}. It sees only one excitation, which apparently becomes overdamped
below T$_{d}$. Only below T$_{c}$ the CM slows down so much that the SM response is no
longer overlapped by the CM and therefore the SM re-appears in INS spectra.

\begin{figure}
\centerline{\includegraphics[width=80mm,clip=true] {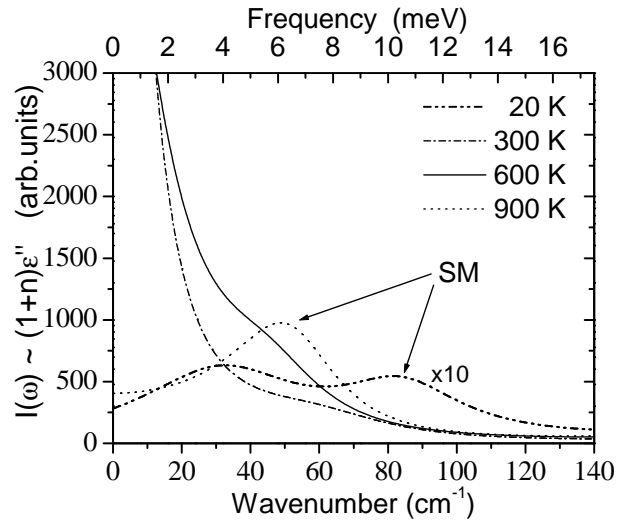}} \caption{Theoretically
calculated inelastic neutron scattering spectra at selected temperatures. The signal is
very small at 20\,K, therefore it was multiplied by 10. Note that experimental INS
spectra are frequently presented only above 3\,meV. \label{fig:neutron}}
\end{figure}

Our suggestions can be supported also by simulation of INS spectra from our
$\varepsilon''(\omega)$ spectra. INS spectra I($\omega$) are proportional to
$(1+n).\varepsilon''(\omega)$, where $n=1/(exp(\hbar\omega/kT)-1)$ is the Bose-Einstein
factor. The convolution with instrumental resolution function of neutron spectrometer
causes additional enhancement of experimental intensity at low ($<$ 3 meV) frequencies,
which is not taken into account. Nevertheless, the simulated
$\varepsilon''(\omega)/\omega$ spectra plotted at several temperatures in Fig.
~\ref{fig:neutron} support our idea. Clear phonon peak is seen near 7 meV at 900\,K, but
this peak is overlapped by the CM at lower temperatures. The TO1 phonon clearly appears
again at 20\,K near 11 meV. Fig.~\ref{fig:neutron} can be compared with experimental
data, e.g. Fig. 3 in Ref. \cite{Wakimoto02a}, where it is clearly seen that the intensity
of the SM in INS spectra is much lower than the total INS intensity at 500 K, where only
the CM was distinguished.

We can also explain why the waterfall effect appears at different wavevectors $q_{WF}$ in
different Brillouin zones\cite{Hlinka03}. The central peak is the most intensive in the
$\Gamma$-point of the Brillouin zone and distinctly losses its intensity with increasing
wavevector. Above $q_{WF}$ the central peak could be so weak that the TO1 phonon branch
appears in the INS spectra. The central peak is in INS spectra remarkably stronger in the
020 Brillouin zone than in the 030 zone, therefore also $q_{WF}$ in the 020 zone is
larger than in the 030 zone.

\section{Conclusion}

It was shown that the phonon behavior is the same in FTIR spectra of PMN single crystal
and thin film. Only CM behavior is responsible for different low-frequency dielectric
response in thin film and single crystal. In contrast to INS data, FTIR spectra of PMN
revealed an underdamped SM at all temperatures between 20 and 900 K. The SM frequency,
which softens only partially, obeys the Cochran law up to 450\,K, then levels off near
50\,\cm\ and slightly hardens above 650\,K. An overdamped CM appears in the FTIR spectra
below the Burns temperature and slows down to MW and lower frequency range on cooling.
Its temperature behavior was directly investigated by high-frequency and MW dielectric
spectroscopy down to 100 K. It is suggested that the waterfall effect in INS spectra can
be explained by overlapping of the SM response with the CM response. A new heavily damped
excitation near 20\,\cm\ appears in the FTIR spectra below $\sim$ 200\,K, probably due to
activation of some short wavelength phonons as a consequence of breaking of translation
symmetry in the disordered PMN.

\begin{acknowledgments}
We are grateful to J. Hlinka for useful discussions, to M. Berta for the help with the
fits of FTIR spectra and to I.P. Bykov for providing the single crystal. This work was
supported by the Grant Agency of Academy of Sciences (projects Nos. A1010203 and
AVOZ1-010-914), Grant Agency of the Czech Republic (projects No. 202/04/0993) and
Ministry of Education of the Czech Republic (project COST OC 525.20/00).
\end{acknowledgments}

\end{document}